\documentclass[letterpaper]{aa}
\usepackage{txfonts,graphicx,epsfig,float,natbib}
\bibpunct{(}{)}{;}{a}{}{,}

\begin{document}

\title{Observations of the companion to the pulsar PSR
  B1718$-$19\thanks{Based on observations collected at the European
  Southern Observatory, Chile (ESO Programme 67.D-0058)}}
\subtitle{The Role of Tidal Circularisation}
   \author{T. Janssen
          \inst{1}
          \and
          M. H. van Kerkwijk\inst{1,2}
          }
   \offprints{T. Janssen}
   \institute{Astronomical Institute, Utrecht University,
              P.O. Box 80000, 3508 TA Utrecht, The Netherlands\\
              \email{T.Janssen@phys.uu.nl}
   \and
             Department of Astronomy and Astrophysics, University of Toronto,
	     60 St George Street, Toronto, Ontario M5S 3H8, Canada\\
	     \email{mhvk@astro.utoronto.ca}
             }
\titlerunning{The companion of PSR B1718$-$19}

\abstract{We present optical and infrared observations taken with the
Very Large Telescope of the eclipsing binary pulsar system \object{PSR
B1718$-$19}. The candidate companion of the pulsar, identified earlier
in {\em Hubble Space Telescope} observations, has been detected in all
three bands, R, I, and J.  These detections allowed us to derive
constraints on temperature, radius, and mass, pointing to a companion
that has expanded to a radius between one of a main sequence star and
one at the Roche-limit.  We focus on the role of tidal circularisation
in the system, which will have transformed the initially eccentric
orbit expected from formation scenarios into the nearly circular orbit
presently observed.  Based on simple energy balance arguments, we are
able to draw a picture of the companion's evolution resulting from the
energy deposition in the star due to circularisation. In this picture,
our measurement of the companion's parameters is consistent with the
expected initial eccentricity.  However, with the present
understanding of tidal dissipation it remains difficult to account for
the short time in which the system was circularised.

\keywords{ binaries: eclipsing --
           pulsars: individual (PSR B1718$-$19) -- 
           Stellar dynamics } }

\maketitle

\section{Introduction}
The binary radio pulsar \object{PSR B1718$-$19} has a 1-s spin period
and was discovered by \citet{1993Natur.361...47L} in the direction of
globular cluster \object{NGC 6342}.  The pulsar is young and has a
high magnetic field: from the pulse period and its derivative one
infers $\tau_{\rm char}=P/(2\dot{P})=10$~Myr and
$B=1.5\times10^{12}$~G.  The binary orbit is circular, has a period of
6.2hr, and is accompanied by eclipses at low radio frequencies, of
400--600 MHz.  The eclipsing material is probably the companion's
wind.

The formation scenarios proposed
\citep{1993Natur.361...47L,1993A&A...273L..38E,1993ApJ...415L.115W,1993MNRAS.264L...3Z,1994ApJ...430L..57B,1996A&A...307..768E}
fall into two main categories: (i) partial recycling of an old neutron
star in a close encounter with other stars in the cluster core, and
(ii) accretion-induced collapse (AIC) of a white dwarf.
\citet{2000ApJ...529..428V} detected a candidate for the pulsar
companion. The object's faintness was a strong indication for scenario
(i), though scenario (ii) could not be excluded definitively.

\citet{1993ApJ...415L.115W} noted that either formation scenario leads
to an eccentric binary orbit, and tidal circularisation must have
worked to reduce the eccentricity to the currently observed low value
($e\lesssim 0.005$). Thus, \object{PSR B1718$-$19} might serve as an
observational constraint on circularisation theory in a parameter
region different from the region covered by constraints that are
currently available: tidal cutoff periods of solar-type binaries (see,
e.g., \citealt{2004ApJ...602L.121M} and references therein), binaries
containing a giant star \citep{1995A&A...296..709V}, or planets
\citep{2003sfre.conf..213W}.

The circularisation efficiency depends strongly on the object's
radius, but this is poorly constrained by the single-band detection of
\citet{2000ApJ...529..428V}.  This motivated us to collect further
observations to obtain a direct measurement of temperature and radius.
In this paper, we describe the results.  In \S\ref{secobs}, we
describe the reduction and show our results in colour-colour and
colour-magnitude diagrammes.  We use these in \S\ref{secana} to put
further constraints on mass, radius and temperature of the companion,
and in \S\ref{secdis} we investigate the implied role of tidal
circularisation, focusing on the circularisation timescale and the
consequences of tidal dissipation in the companion.

\section{Observations\label{secobs}}

\begin{table*}
\caption{Log of observations\label{obstab}}
\begin{minipage}[t]{\textwidth}
\renewcommand{\thefootnote}{\thempfootnote}
\renewcommand{\footnoterule}{}
\begin{center}
\begin{tabular}{cccccccc}
\hline
Instrument & Band & Night & $t_{\rm{int}}$ (s)    & seeing (\arcsec) & \# EXP    & STD-field\footnote{The standard magnitudes are taken from \citet{2000PASP..112..925S} for I and R and from  \citet{1998AJ....116.2475P} for J.}   & $\Delta \sec z$\footnote{This column lists the difference in airmass between the single exposure science frame that was used for calibration and the standard frame (see text).}\\
\hline\\[-5pt]
VLT-FORS2 & I & 2001 Jun 20 & 257 & 0.48-0.61 & 8 & PG1323 & 0.04\\
& & 2001 Jun 22 & 240, 257 & 0.44-0.67 & 8 & SA92-249 & 0.00\\[5pt]

VLT-FORS2 & R & 2001 Jun 20 & 317 & 0.51-0.68 / 0.75\footnote{The frames were selected on seeing value. Listed in the seeing column is, before the slash, the seeing range of frames included in the mosaic image, and, after the slash, the seeing range of frames that were excluded. In the \# EXP column the number of exposures included in the mosaic image is displayed, with after the slash the total number of exposures taken.} & 11 / 12\footnotemark[\value{mpfootnote}] & PG1323 & 0.02\\
& & 2001 Jun 22 & 300, 317 & 0.48-0.66 / 0.71-0.83\footnotemark[\value{mpfootnote}] & 11 / 13\footnotemark[\value{mpfootnote}] & SA92-249 & 0.02\\[5pt]

VLT-ISAAC & J & 2001 Jun 7  & 30.30 & \ldots\ / 0.59-0.87\footnotemark[\value{mpfootnote}] & 0 / 16\footnotemark[\value{mpfootnote}] & S860-D & 0.00\\
& & 2001 Jun 14 & 30.30 & 0.36-0.56 & 48 & S860-D  & 0.07\\[5pt]

$HST$-WFPC2 & F702W & 1997 Mar 8 & 700 & 0.046 & 12 & \ldots & \ldots\\
\hline
\end{tabular}
\end{center}
\end{minipage}
\end{table*}

The observations on which we report here were done for us in Service
Mode in June 2001 with the Very Large Telescope (VLT) at ESO (Paranal)
through filters R, I and J (see Table~\ref{obstab}).  In this section
we will describe their reduction and photometry.

We will also use the data presented in \citet{2000ApJ...529..428V}:
images obtained through the F702W filter with the Wide Field and
Planetary Camera 2 (WFPC2) of the {\em Hubble Space Telescope}
(\emph{HST}) in March 1997.  Only the data from the Planetary Camera
(800$\times$800 pixels, at $0\farcs046~{\rm pixel}^{-1}$) will be used
for the present paper.  As will be discussed in \S\ref{hstprob}, we
were unable to derive reliable photometry from these data. The F702W
images still serve, however, for measurements of the proper motions of
stars in the field.

\begin{figure*}[t]
\sidecaption
\includegraphics[bb=72 345 502 745,width=12cm,clip=]{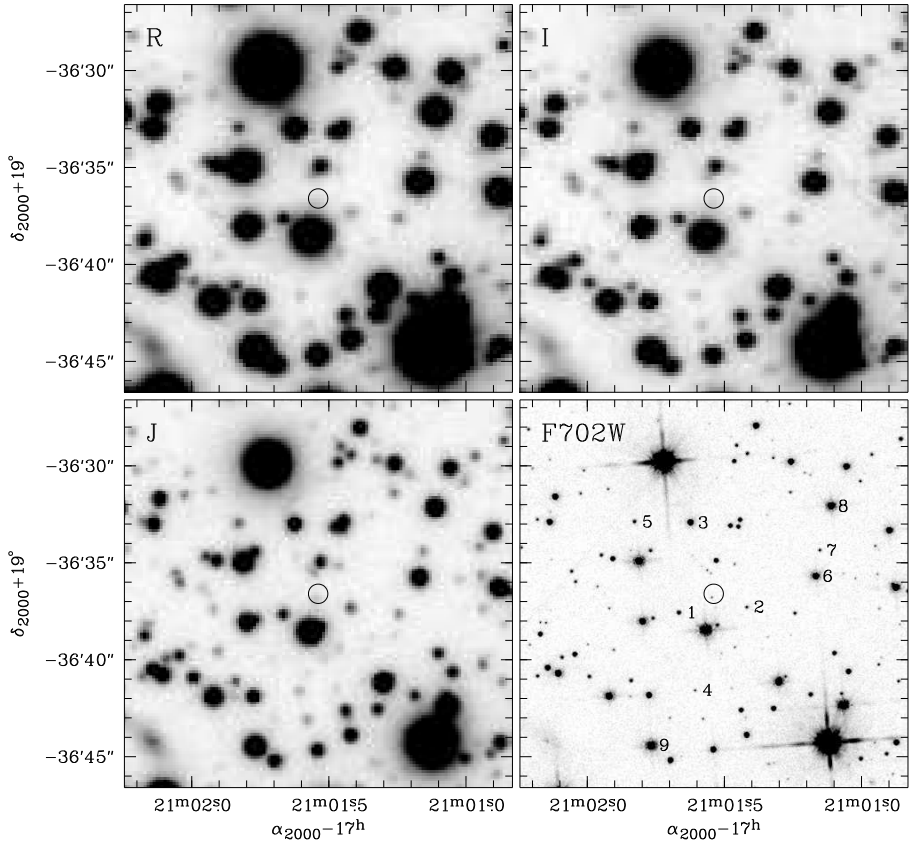}
\caption{Images of the field of \object{PSR B1718$-$19}. Shown are --
clockwise -- the images in R and I (taken with VLT-FORS2), in F702W
($HST$-PC), and in J (VLT-ISAAC). The F702W-observations have been
presented in \citet{2000ApJ...529..428V}, who also performed the
astrometry to derive the 95\% confidence error circle of radius
0\farcs5, centered on the position of the pulsar.  This error circle is
displayed in the images, and contains a detection of the candidate
counterpart in all bands. For stars indicated by numbers in the
F702W-panel, and for the PSR-companion we list positions and
magnitudes in Table~\ref{appmag}.
 \label{ima}}
\end{figure*}

The astrometry done by \citet{2000ApJ...529..428V} enables us directly
to identify the candidate for the pulsar's companion in our images,
shown in Fig.~\ref{ima}.

\subsection{Infrared observations in J}

The J-band observations were done with the short-wavelength channel of
the Infrared Spectrometer And Array Camera
\citep[ISAAC;][]{1998Msngr..94....7M}, mounted on the 8.2m Antu (Unit
Telescope 1).  The detector was a Rockwell Hawaii 1024x1024 pixels
array, with a plate scale of $0\farcs148~{\rm pixel}^{-1}$.

The ISAAC-observations need to be corrected for two instrumental
effects: the \emph{odd-even column effect}, which can be seen as an
offset between the odd and even columns of the array, and the
\emph{electrical ghost}, which consists of an additional signal
proportional to the sum of the intensity along a given row \emph{and}
the row 512 rows away \citep[see the ISAAC Data Reduction
Guide\footnote{Available at
http://www.eso.org/instruments/isaac/drg/drg.ps},][]{isaac-drg}.  We
employ ESO's software package {\sc eclipse} \citep{eclipse} for these
corrections.  Next, we use {\sc midas} to subtract the dark current
from all frames and divide them by twilight flat fields.

From the images, it was clear that the seeing in the first observing
night was much worse than in the second night (see
Table~\ref{obstab}).  We decided to exclude all 16 frames taken on the
first night. We register the 48 frames of the second night with
respect to each other, applying offsets rounded to the nearest integer
pixel, and compute an average image. Bad pixels are excluded from this
average.  We regard a pixel as bad, if (i) it is located at rows 513
or 514 (2048 pixels), where bias-level variations with time prevent
proper bias subtraction; (ii) the difference between a long and a
short dark-current exposure at that location lies more than
7.5$\sigma$ away from the mean (57 pixels); (iii) it has a value
lower than $2.75\sigma$ below background in more than 50\% of the
frames (172 pixels), or (iv) if it is a hot pixel (59 pixels).
Finally, in order to prevent problems with differences in the exposure
time and distortion near the edges, we extract the central
$600\times600$ pixels ($89\times89\arcsec$) part of our mosaic image.
In this subimage, the point-spread function (PSF) shows negligible
variation.

We perform PSF-fitting photometry using the DAOPHOT II package
\citep{1987PASP...99..191S}.  The determination of the PSF is hampered
by the crowdedness of the field: one object per 4 arcsec$^2$, at a
full width at half maximum (FWHM) of 0\farcs5.  We deal with this
following the suggestions of \citet{1987PASP...99..191S}, by first
estimating the PSF based on several relatively isolated stars, and
then using this PSF to subtract neighbouring stars.  On the new image
a better PSF can be determined.  This procedure has been repeated four
times.  With the resulting final PSF, the magnitudes of all stars are
determined.

\begin{table}[t] 
\begin{center} 
\caption{Positions and apparent magnitudes of the candidate companion
of \object{PSR B1718$-$19} and surrounding stars.  The uncertainty in
the photometric zero point is 0.02~mag for J and $<\!0.03~$mag in R \&
I.  The uncertainty in the astrometric tie is $0\farcs 06$ in each
coordinate \citep{2000ApJ...529..428V}.  Photometry for all stars in
the field, including uncertainties, is available online.
\label{appmag}} 
\begin{tabular}{ccccccc} 
\hline 
ID & $\alpha_{2000}$ & $\delta_{2000}$  & $R$ & $I$ & $J$ \\ 
& 17$^{\rm h}$21$^{\rm m}$ & -19\degr36\arcmin\\
\hline
 PSR&$01 \fs 548$ & $36 \farcs 77$ &    25.87 &    24.25 &    22.59\\
&$\pm0 \fs 004$ & $\pm0 \farcs 06$&$\pm 0.20$&$\pm 0.08$&$\pm 0.08$\\
   1&$01 \fs 667$ & $37 \farcs 55$ &    23.31 &    22.09 &    20.56\\
   2&$01 \fs 421$ & $37 \farcs 26$ &    24.79 &    22.98 &    21.08\\
   3&$01 \fs 625$ & $32 \farcs 91$ &    21.00 &    20.35 &    19.37\\
   4&$01 \fs 608$ & $41 \farcs 54$ &    25.74 &    23.96 &    22.03\\
   5&$01 \fs 828$ & $32 \farcs 87$ &    23.55 &    22.53 &    21.14\\
   6&$01 \fs 171$ & $35 \farcs 67$ &    20.50 &    19.82 &    18.75\\
   7&$01 \fs 156$ & $34 \farcs 34$ &    24.82 &    23.49 &    21.67\\
   8&$01 \fs 114$ & $32 \farcs 06$ &    20.02 &    19.40 &    18.49\\
   9&$01 \fs 767$ & $44 \farcs 39$ &    19.59 &    18.95 &    17.99\\
\hline 
\end{tabular} 
\end{center} 
\end{table} 

Observations of a standard star (Table~\ref{obstab}) serve to
calibrate the photometry.  As these observations are taken with very
short exposure times, the PSF varies greatly over the field.  Hence,
PSF-fitting is not reliable and we resort to aperture photometry. The
standard star's aperture magnitudes can then be related to magnitudes
of stars in a single exposure science frame, measured with a
corresponding aperture.  For this purpose, we select a science frame
taken at airmass as close as possible to that at which the standard
star was taken (see Table~\ref{obstab}), so that there is no need to
correct for atmospheric extinction.  To account for the quite
different seeing of the standard and the science exposures, we use
apertures with an equal radius in units of the image FWHM, not
pixels.  We used the trend of magnitude as a function of aperture to
verify that using an aperture of the same radius in FWHM indeed
results in a superior match of the fraction of the flux inside a given
aperture in two different frames.

Next, we compute an aperture correction to relate the aperture
magnitude in the single science frame to the magnitudes determined
from PSF fitting in the mosaic image.  Out of necessity, since we do
not have another infrared colour, we neglect any colour terms.  From
the ISAAC Data Reduction Guide \citep[\S 5.5.1]{isaac-drg}, this
appears reasonable: the $J-K$ colour term is expected to differ from 0
by less than 0.01.  Overall, we estimate that the total error in the
calibration is $\sim0.02$~mag.

\subsection{Optical observations in R and I}
The R and I-band observations were carried out with the FOcal
Reducer/low dispersion Spectrograph
\citep[FORS2;][]{2000SPIE.4008...96S} mounted on Yepun (VLT-UT4).
FORS was used in Standard Resolution imaging mode, in which the plate
scale on the $2048\times2048$ pixels SITe detector is $0\farcs201~{\rm
pixel}^{-1}$.

The reduction was done in a similar fashion as described above for the
ISAAC-observations.  First, all frames are bias-subtracted and divided
by twilight flat fields.  We select frames with a seeing lower than
$0\farcs7$ (see Table~\ref{obstab}) and compute an average mosaic of
the registered frames, excluding 1329 pixels in each image located on
a bad column, as well as pixels that were hit by cosmic rays.  From
the mosaics, we extract the central $400\times400$ pixels
($80\times80\arcsec$) parts, which are covered by all individual
frames and which have negligible distortion.  On these subimages the
photometry is done using procedures described above for the ISAAC
data.  For the calibration, we determine a weighted average from both
observing nights for the zero point and the color term.  This leads to
a total error in the calibration of $\la\!0.03~$mag in both bands.

\subsection{Problems with the F702W-observations}\label{hstprob}

We tried to obtain photometry from the {\em HST} F702W observations,
both using {\sc HSTphot} \citep{2000PASP..112.1383D} and following the
instructions in the {\em HST} WFPC2 data handbook \citep{hstdata}.
We failed to obtain numbers that we felt were sufficiently reliable,
for two reasons.  First, indepent of the reduction method, we found
that in colour-colour diagrammes (e.g., $m_{\rm F702W}-I$ vs. $I-J$)
the root-mean-square scatter for bright, blue stars ($I-J<1.05$) was
as large as 0.026~mag, while in a comparable plot for $R-I$ (see
Fig.~\ref{cmdccd}), it is only 0.008~mag.  This implies the WFPC2
data had a base noise of $\sim\!0.02$ to 0.03~mag, which we do not
understand.  From discussions with Drs Andrew Dolphin and Peter
Stetson (2004, personal comm.), we learned that such scatter is
normally present, but that it should average out when one uses
multiple frames.  In our case, however, it did not, possibly because
our dither pattern used integer pixel offsets.  We also confirm it is
present in other comparisons between ground-based and WFPC2 data (such
as those presented in \citet{2001A&A...378..986V}; for that field, we
find that ACS/HRC data show much less scatter and are completely
consistent with the ground-based data).

The second and more severe problem is that the two reduction schemes
lead to a zero point that differs by 0.1~mag.  We are confident that
this is not because we did an inappropriate aperture correction (such
as the 0.1~mag offset between the standard aperture of $0\farcs5$ and
``nominal infinity''); we do not find such differences for images in
other filters that we analysed.  Since we were unable to identify the
flaw in the reduction, we conclude that we cannot rely on the
magnitude derived for the pulsar companion, and we excluded the F702W
photometry from our analysis.  We note, however, that within a 0.1~mag
uncertainty, the magnitude found is consistent with that expected from
our R and I-band detections.

\subsection{Photometry and Proper Motions}\label{secpmcmdccd}

\begin{figure}
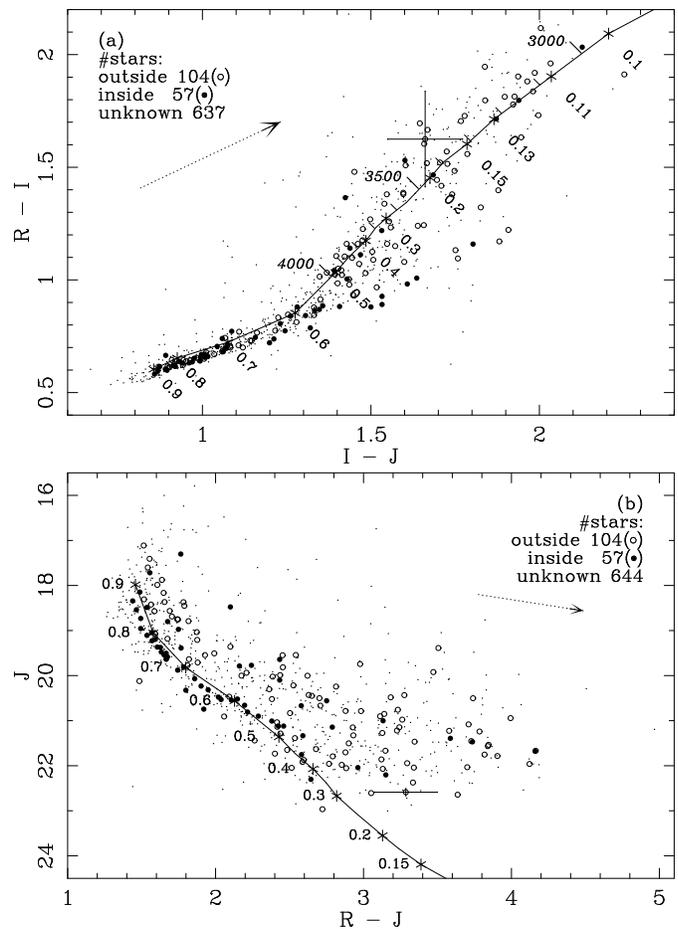

\begin{center}
\resizebox{\hsize}{!}{\includegraphics[height=\columnwidth,angle=270]{teximages/arthrd_RsubI_IsubJ.ps}}\\[2pt]
\resizebox{\hsize}{!}{\includegraphics[height=\columnwidth,angle=270]{teximages/arthrd_J_RsubJ.ps}}
\end{center}
\caption{\textbf{a} Colour-colour and \textbf{b} colour-magnitude
diagrammes of the companion of PSR B1718$-$19 (error bars) and stars in
the field.  The filled circles denote stars with a proper motion
inside the dashed circle of Fig.~\ref{figpm}, which are likely
associated with the cluster NGC 6342.  The open circles indicate stars
with a proper motion outside this circle.  For stars denoted by a dot,
no proper motion is available, since they were outside the field of
view of the PC.  The detection limit in J is around $23\,$mag.  The
solid line represents model main sequence stars, labeled with their
masses (in solar units), and -- in italic typeface -- temperature
(small tickmarks denote intervals of 100~K).  The models are corrected
for a reddening $E_{V-I}$=0.55 and distance modulus $(m-M)_0$= $14.69$
(see \S\ref{secana}).  The dotted arrow denotes a reddening of
$E_{V-I}$=$0.55 \pm 0.07$ (with the size of the solid part equal to
the uncertainty).
\label{cmdccd}}
\end{figure}

We used our calibrated photometry (Table~\ref{appmag} and on-line
material), to construct colour-colour and colour-magnitude diagrammes;
see Fig.~\ref{cmdccd}.  In the colour-magnitude diagramme, we find a
rather broad main sequence, which suggests that many of the stars do
not belong to the cluster.

\begin{figure}
\begin{center}
\includegraphics[height=\columnwidth,angle=270]{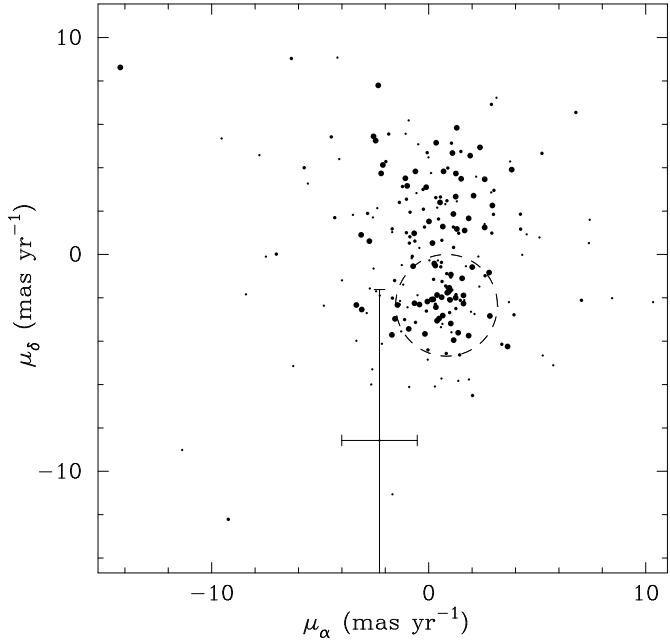}
\end{center}
\caption{Proper motion of the counterpart of PSR B1718$-$19 (error
bar) and stars in the field.  The three different sizes of the dots
correspond to three brightness classes,: $J<20.3$, $20.3\le J\le21.7$,
and $J>21.7$.  The dashed circle, with a radius of 2.3 mas yr$^{-1}$,
indicates the group of stars which we associate with the cluster.  The
counterpart of PSR~B1718$-$19 has a proper motion consistent with our
estimate of the cluster's proper motion (the uncertainty in
declination is much larger than that in right ascension due to the
presence of a bright star to the south of the companion).
\label{figpm}}
\end{figure}

As the {\em HST} observations were taken 4.25 years prior to the
VLT-observations, we can try to separate cluster stars from fore- and
background objects using proper motions.  For this purpose, we
measured positions in the {\em HST} PC and our best-seeing VLT
observation, that in the J band.  The result is shown in
Fig.~\ref{figpm}.  One sees that especially among brighter stars, a
clustering of stars associated with NGC~6342 is apparent.  This is
confirmed in Fig.~\ref{cmdccd}, where we use filled symbols to
indicate stars that are likely cluster members (those within the
dashed circle in Fig.~\ref{figpm}): the brighter ones, with
$J\lesssim21$, form a clear, narrow sequence that is well matched by
the expected main sequence for NGC~6342.

Nearly all bright stars that have proper motions inconsistent with
cluster membership (outside the dashed circle; open symbols in
Fig.~\ref{cmdccd}) are brighter at a given colour than the
cluster-associated ones. This is expected, since at the galactic
latitude of $9\fdg73$ of our observations, most stars belonging to the
disk should be in front of the cluster.  We can estimate a typical
distance by considering the number of stars per square arcsecond at a
given distance.  For an exponential distribution, this should scale as
$n(d) \propto d^2\exp(-d sin|b|/z_0)$, which reaches a maximum at
$d_{\rm max} = 2 z_0/sin|b|$.  Given a latitude $b=9\fdg73$ for NGC
6342 \citep[][February 2003 revision]{1996AJ....112.1487H} and scale
height $z_0=325$~pc \citep{1993ApJ...409..635R}, we obtain $d_{\rm
max}\simeq3.8$~kpc.  Since the distance of NGC 6342 is 8.6~kpc, we
would thus expect stars in the galactic disks to be typically
$\sim\!1.7$~mag brighter than stars in NGC 6342.  This is roughly
consistent with the offset seen in Fig.~\ref{cmdccd}b.

From Fig.~\ref{figpm}, one sees that for stars with $J\gtrsim21$, the
distinction between cluster-associated and foreground stars is much
less clear.  This is due to the increasing uncertainty in the J-band
position (which dominates the proper motion uncertainty).  For stars
with $J\lesssim20$, the position's uncertainty is $\sim\!2$~mas, while
around $J\approx21$, it has increased to $\sim\!4$~mas, reaching up to
$\sim\!10$~mas for stars at the detection limit ($J\approx23$).  The
latter error corresponds to an error in the proper motion of
2.3~mas~yr$^{-1}$, equal to the radius of the dashed circle in
Fig.~\ref{figpm}.

\section{Analysis\label{secana}}

In this section, we estimate the companion's radius and temperature
from its observed magnitudes.  Since it is associated with NGC 6342,
we can use the distance modulus $(m-M)_0$= $14.69 \pm 0.20$ found by
\citet{1999A&A...347..455H}.  \citeauthor{1999A&A...347..455H} also
determined the reddening, but found a rather large differential
variation.  Below, therefore, we first determine the reddening in our
field, by fitting models to cluster stars in the colour-colour and
colour-magnitude diagrammes.  

Both to determine the reddening and to infer the companion's
properties, we use the evolutionary tracks for low-mass stars from
\citet{1998A&A...337..403B} (hereafter, BCAH models).  We use the
tracks for metallicity ${\rm\lbrack M/H\rbrack}=-0.5$, which
corresponds to ${\rm[Fe/H]}\simeq-0.77$ \citep{1997A&A...327.1054B},
close to the measured metallicity ${\rm\lbrack Fe/H\rbrack}=-0.65$ for
NGC 6342 (\citealt{1996AJ....112.1487H}, February 2003 revision).

\subsection{Reddening\label{secanared}}

\citet{1999A&A...347..455H} found strong evidence for differential
reddening for NGC~6342, with $E_{V-I}$ varying between 0.64 and 0.74
in the core region of the cluster.  At the location of PSR~B1718$-$19,
they inferred $E_{V-I}=0.70\ldots0.76$.

In Fig.~\ref{cmdccd}a, we see that the colours of the likely cluster
members are at best marginally consistent with such high reddening; a
much better fit is found for a value $E_{V-I}\simeq0.55$.  To try to
understand this discrepancy, we tried to determine whether the
reddening varied over our smaller, $80\times80\arcsec$~field.  We find
some evidence for variation, with $E_{V-I}\simeq0.40$ at the south
end, and $E_{V-I}\simeq0.75$ at the north end.  Thus, it may be that
the value found by \citet{1999A&A...347..455H} simply reflects a
different effective centre of the region in which the reddening was
determined.

Looking in detail at Fig.~\ref{cmdccd}, one sees that for masses above
$\sim\!0.6\,M_\odot$, the models are slightly too red in $R-I$.  This
could be corrected by decreasing the reddening further, but then the
break in the curve at $\sim\!0.6\,M_\odot$ would happen at an $I-J$
colour that is too blue.  The fact that we cannot obtain a better
overall fit is likely related to inaccuracies in the models
\citep[see][]{astroph..0406215}.  To take this uncertainty into
account, we adopt a relatively large uncertainty in $E_{V-I}$, of
0.07~mag.  

We conclude that the reddening to PSR B1718$-$19 is
$E_{V-I}=0.55\pm0.07$ (or $E_{B-V}=0.40\pm0.05$).  Converting
reddening to extinctions in the various bands
(\citealt{1998ApJ...500..525S}, Table 6), we find $A_R=1.07$,
$A_I=0.78$, and $A_J=0.36$.

\subsection{Radius \& Temperature\label{secana_rt}}

Below we will determine the radius and temperature through a direct
fit to the models.  Before doing so, we will make a cruder but more
insightful estimate.  We looked at the BCAH-models and found that, to
satisfying approximation, colors depend on temperature and not on
surface gravity, as long as $\Delta\log g\lesssim1$.

From Fig.~\ref{cmdccd}a, we see that the companion's colours are
consistent with those of main sequence stars with a temperature of
$\sim\!3400\pm150$~K.  The influence of the error in the reddening can
be estimated by moving the star over the solid part of the reddening
vector; this induces an error of $\sim\!70$~K.  Adding this in
quadrature, the final estimate of the temperature is $3400\pm170$~K.

The companion's radius can now be estimated from Fig.~\ref{cmdccd}b.
The masses of main-sequence stars with temperatures consistent with
that of the companion lie in the range $0.13\lesssim M \lesssim 0.27
M_\odot$.  Figure~\ref{cmdccd}b shows that these stars are fainter
than the companion, which implies that the companion has a larger
radius, by a factor $10^{-0.2\Delta J}$ (where $\Delta J$ is the
brightness difference).  From Fig.~\ref{cmdccd}b, we estimate $\Delta
J\simeq0.3$ mag and $\Delta J\simeq2.0$ mag for MS-stars of mass 0.27
and $0.13~M_\odot$, respectively.  Given the main-sequence radii from
the models (stars with mass $(0.13,0.25,0.30)~M_\odot$ have radii
$(0.15,0.26,0.30)~R_\odot$) one infers a radius of
$\sim\!0.31~R_\odot$ at the high temperature end, and $0.37~R_\odot$
at the low end.

For a given temperature, the uncertainty in the radius thus derived
has a contribution from the uncertainties in the observations and the
reddening (about 0.09~mag for the J band), but it is dominated by the
0.2~mag uncertainty in the distance modulus.  The uncertainty in the
radius for given temperature is therefore $\sim\!10$\%. 

\begin{figure}[t]
\begin{center}
\includegraphics[height=\columnwidth,angle=270]{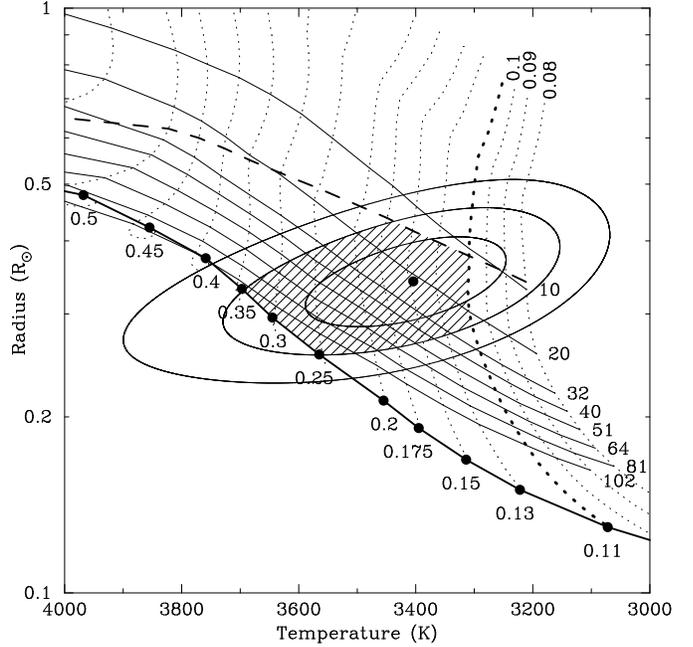}%
\end{center}
\caption{Constraints on the radius and temperature of the companion of
PSR~B1718$-$19.  The ellipses show the 1, 2, and 3-$\sigma$ contour
levels ($\Delta\chi^2=2.30$, 6.17, and 11.8).  The \textbf{bold
continuous line} connecting the filled circles labeled with stellar
masses (in solar units) indicates the model main sequence stars (the
BCAH-models with age 10 Gyrs). The \textbf{dotted lines} connect
BCAH-models in a pre-main-sequence (PMS) contraction phase; the
\textbf{bold dotted line} marks the lower limit set to the companion
mass by the pulsar mass function.  These PMS-tracks are close to the
Hayashi-limit, and stars in hydrostatic equilibrium can be only on or
to the left of them.  The \textbf{bold dashed line} connects the
Roche-radii for stars on the PMS-tracks.  The hatched region shows the
full range of allowed parameters in the $2\sigma$ confidence region
(see Table~\ref{constab}).  Finally, the \textbf{thin continuous
lines} are isochrones labeled with their age in Myrs.  These
isochrones show that, within the system age of 10 Myrs, the companion
cannot have shrunk back to the main sequence if any significant
bloating occurred that involved the whole star.}
\label{contour}
\end{figure}

Now that we know what temperature, radius, and associated
uncertainties we expect from our observations, we proceed with a fit.
We want to take into account the uncertainty in distance and reddening
and hence define $\chi^2$ as
\begin{eqnarray}\label{chi2}
\chi^2 &=& \sum_{\rm X} 
  \left(\frac{m_{\rm X} - A_{\rm X}^{\rm var} - \Delta m_{\rm var} 
              + 5 \log\left(R_{\rm var}/R_{\rm mod}\right) 
              - M_{\rm X}^{\rm mod}}
             {\sigma_{\rm X}}
  \right)^2 \nonumber\\ &&
+ \left(\frac{E_{V-I}^{\rm var} - E_{V-I}^{\rm lit}}
             {\sigma_{E,\rm lit}}
  \right)^2 
+ \left(\frac{\Delta m_{\rm var} - \Delta m_{\rm lit}}
             {\sigma_{\Delta m,\rm lit}}
  \right)^2
\end{eqnarray}
with $A_{\rm X}^{\rm var}=(A_{\rm X}/E_{V-I})E_{V-I}^{\rm var}$, and
$\Delta m = (m-M)_0$.  The label "var" indicates that that part of the
equation can be varied to minimize $\chi^2$, and the symbol X denotes
the filter:~R, I, or J.  Thus, we have five measurements (three
magnitudes $m_R$, $m_I$, and $m_J$, the distance modulus $\Delta
m_{\rm lit}$, and reddening $E_{V-I}^{\rm lit}$) and four parameters
(radius and temperature [which enter the model magnitudes], true
distance modulus $\Delta m_{\rm var}$ and reddening $E_{V-I}^{\rm
var}$).  Using Eq.~\ref{chi2}, we create a contourplot of $\chi^2$ as
a function of radius and temperature in Fig.~\ref{contour}.  Here, we
have marginalized over reddening and distance modules (i.e., for every
radius and temperature, reddening and distance modules are adjusted to
minimize $\chi^2$).  We see that the result is consistent with the
crude estimate made above.

From Fig.~\ref{contour}, we see that the companion likely has a radius
larger than that of a main-sequence star of the same mass.  The
companion mass is constrained by the pulsar mass function ($M_{\rm
C}\gtrsim 0.11~M_\odot$ for a $1.35~M_\odot$ pulsar,
\citealt{1993Natur.361...47L}).  Furthermore, for given mass, the
radius of the companion has to be smaller than the Roche lobe, and the
temperature cannot be cooler than the Hayashi limit, where a star is
completely convective.  The latter limit implies that for given mass,
the companion should be on or hotward of the corresponding
pre-main-sequence track.

The range of allowed parameters of the candidate companion is listed
in Table~\ref{constab}.  At the 1-$\sigma$ confidence level, we
observe a star with $0.11\lesssim M \lesssim 0.27$ M$_\odot$ that is
bloated compared to a main-sequence star of the same mass.  Its radius
lies below the Roche-limit for the mass range $M \gtrsim
0.14~M_\odot$.  However, at the 2-$\sigma$ level we cannot exclude
either a somewhat more massive main-sequence star or a lower-mass,
Roche-lobe filling star (see Fig.~\ref{contour}).  As we will see
below, however, our most likely intermediate position may be most
consistent with expectations from the preceding history.

\begin{table*}
\caption{Companion parameters, at $2\sigma$ confidence level and at
maximum probability, and corresponding circularisation
timescales.\label{constab}}
\begin{minipage}[t]{\textwidth}
\renewcommand{\thefootnote}{\thempfootnote}
\renewcommand{\footnoterule}{}
\begin{center}
\begin{tabular}{lccccc}
\hline
Position in & Temperature & Radius & Mass\footnote{The mass of the companion is estimated in Fig.~\ref{contour} through interpolation of the BCAH-models.} & $\tau_{\rm{circ}}$ $[Z]$\footnote{The circularisation timescales are computed in two prescriptions. One is described in \citet[indicated with {$[Z]$}]{1989A&A...220..112Z}, the other in \citet[indicated with {$[G\&O]$}]{1997ApJ...486..403G}.} & $\tau_{\rm{circ}}$ $[G\&O]$\footnotemark[\value{mpfootnote}]\\
$R,T$-diagram & (K) & (R$_\odot$) & (M$_\odot$) & (Myr) & (Myr)\\
\hline
Main sequence & $3560 - 3700$ & $0.26 - 0.33$ & $0.25 - 0.35$ & $17 - 5.3$ & $(10 - 3.4) \times 10^3$ \\
Maximum probability & 3400 & $ 0.34$ & $0.15$ & $0.41$ & $230$ \\
Roche-lobe & $3310 - 3440$ & $0.38 - 0.43$ & $0.11 - 0.17$ & $0.088 - 0.089$ & $46 - 50$ \\
\hline
\end{tabular}
\end{center}
\end{minipage}
\end{table*}

\section{Discussion\label{secdis}}

We now turn to the implications of our measurements.  In doing this,
we will first assume that the system formed through the scenario of a
three-body interaction in the core of the cluster.  The other, in our
opinion less likely scenario of accretion-induced collapse of a white
dwarf will be discussed seperately in \S\ref{aic}.

If the system formed in a three-body interaction, a bloated companion
is expected.  This is because the binary that is produced is expected
to be highly eccentric, with $e\simeq0.7$
\citep[\S5]{1992RSPTA.341...39P}, while the system's current
eccentricity is very low ($e\lesssim0.005$).  Thus, the orbit must
have circularised within the system age, which is 10~Myr at most
\citep{2000ApJ...529..428V}.  The natural mechanism -- tidal
circularisation -- transfers orbital energy to the companion, thereby
bloating the star.

Below, we first estimate the expected bloating, and then look into
whether tidal theory can account for the observed limit on the
circularisation timescale.

\subsection{Energy balance during circularisation\label{ebal}}

The response of a star (or planet) to the dissipation of tidal energy
has been investigated by a number of authors
\citep[e.g.][]{1987A&A...184..164R,1996MNRAS.279.1104P,2004ApJ...608.1076G,2004MNRAS.347..437I},
but none of these studies is directly applicable to our situation.
Specifically for \object{PSR B1718$-$19}, \citet{1994A&A...285L..21V}
showed that the tidal energy is of the order of the potential energy
of the star, assuming $a_{\rm ini} \gg a_{\rm c}$ or $e_{\rm
ini}\sim1$ (here and below, the labels ``ini'' and ``c'' refer to the
situations where the pulsar binary has just been formed and where the
orbit has been circularised, respectively).  With our constraints on
the companion's radius and temperature, we can make a more precise
picture of the exchange of energy between orbit and stellar interior.

For our estimate, we assume that at the onset of circularisation, the
companion is an ordinary main sequence star.  Given the range of
masses allowed for the companion (see Table~\ref{constab} and
Fig.~\ref{contour}), it will be fully convective.  A measure for the
timescale on which energy is distributed across the star is the
convective friction time $t_{\rm f} = (m R^2 / L)^{1/3}$, which varies
from $0.55$~yr at the high mass end of allowed companion parameters to
$0.43$~yr at the low mass end.  This is much shorter than the thermal
timescale in most of the star and than the circularisation timescale.
We thus assume that the energy dissipated during circularisation is
distributed through the entire star.

In reaction to the increase in internal energy, the star will
expand until hydrostatic and approximate thermal equilibrium is
regained.
For a completely convective star, the total energy $E_\star$ is given by, 
\begin{equation} \label{epot}
E_{\star}=\frac{1}{2}E_{\rm pot}\simeq-\frac{3Gm^2}{7R},
\end{equation}
where we used the Virial theorem and assumed the star's potential
energy $E_{\rm pot}$ was roughly that of a $n=1.5$ polytrope
(\citealt{1939isss.book.....C}, Eq.~90), as appropriate for a
completely convective star.  The symbols $m$ and $R$ denote the
companion's mass and radius, respectively.

Next, we use that orbital angular momentum should be roughly conserved
in the circularisation process (the companion's rotational angular
momentum is much smaller than the orbital angular momentum).  This
implies $a_{\rm ini}(1-e_{\rm ini}^2)=a_{\rm c}(1-e_{\rm c}^2)\approx
a_{\rm c}$, and hence,
\begin{equation} \label{eorb}
E_{\rm orbit, ini} = \frac{GmM}{2a_{\rm ini}}
                   \approx E_{\rm orbit, c}(1-e_{\rm ini}^2),
\end{equation}
where $M$ is the mass of the neutron star, for which we take $1.35
M_\odot$ \citep{1999ApJ...512..288T}, and $a$ the semi-major axis,
which can be computed for given companion mass using the observed
period of 22314.83~s \citep{1993Natur.361...47L}.  Combining
Eqs.~\ref{epot} and \ref{eorb}, we can write the exchange of orbital
energy and total energy of the companion, $\Delta E_{\rm orbit}=\Delta
E_{\star}$, as:
\begin{equation} \label{eexchange}
\frac{GmM}{2a_{\rm c}}e_{\rm ini}^2 = 
  -\frac{3Gm^2}{7} \left[\frac{1}{R_{\rm c}} - \frac{1}{R_{\rm MS}}\right],
\end{equation}
where we used the assumption that the initial radius $R_{\rm
ini}$ is that of a main-sequence star, $R_{\rm MS}$.  Here,
we neglected the contribution from the companion's rotational energy,
since this is much smaller than the change in orbital energy.

\begin{figure}[t]
\begin{center}
\includegraphics[height=\columnwidth,angle=270]{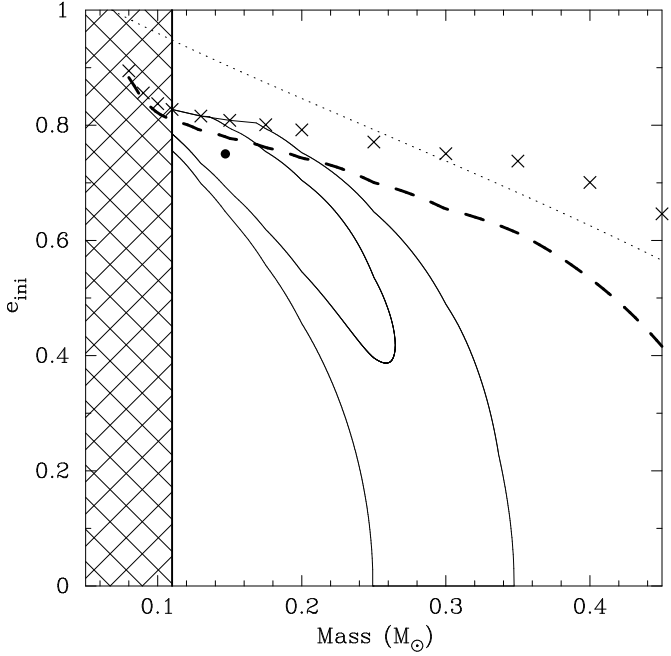}%
\end{center}
\caption{Initial eccentricity e$_{\rm ini}$ for which the change in
orbital energy $\Delta E_{\rm orbit}$ during circularisation equals
the excess energy $E_{\star \rm,obs} - E_{\star\rm,MS}$ of the present
companion relative to a main-sequence star (see Eq.~\ref{eexchange}).
Shown are the initial eccentricities that can account for the bloating
observed at the best-fit point as well as at the 1 and 2-$\sigma$
contours from Fig.~\ref{contour}.  The hatched region indicates the
lower limit to the companion mass of $0.11~M_\odot$.  The dotted line
shows the estimate from Eq.~\ref{ecczero} for the \emph{critical}
initial eccentricity at which the star would be bloated to the Roche
lobe.  The crosses mark the critical eccentricities found from a more
careful calculation; the main difference is due to the use of the more
accurate main-sequence radii from the BCAH-models.  The bold dashed
line indicates the critical eccentricity above which Roche lobe
overflow occurs at any time during the circularisation process (see
text).}
\label{eccini}
\end{figure}

For a given companion mass, we can use Eq.~\ref{eexchange} to
calculate the initial eccentricity that induces bloating of the main
sequence star up to a given radius.  In this way, we convert the
$1\sigma$ and $2\sigma$ contours in the radius-temperature diagramme
(Fig.~\ref{contour}) into corresponding contours in an
eccentricity-mass diagramme (Fig.~\ref{eccini}), where we determine
the mass by assuming the star is presently on a cooling track.

From Eq.~\ref{eexchange} we can also calculate the \emph{critical}
initial eccentricity, which is the eccentricity for which the main
sequence star is bloated exactly to the Roche limit.  For a rough
estimate, we approximate the semi-major axis, which is only weakly
dependent on $m$, by $a_{\rm c}\approx2R_\odot$, and the Roche-limit
by $R_L\approx0.46a(m/(m+M))^{1/3}$ \citep{1971ARA&A...9..183P}.
Applying the mass-radius relation \(R_{\rm MS}/R_\odot \approx m_{\rm
MS}/M_\odot \), we obtain
\begin{equation} \label{ecczero}
e_{\rm crit}^2 \simeq 1.270 - 0.553 \left(\frac{m}{0.2M_\odot}\right)^{2/3}.
\end{equation}
In Fig.~\ref{eccini}, one sees that the approximation agrees rather
well with the values determined directly from the model; the
differences are largely due to the inaccuracy in the assumed
mass-radius relation.

In principle, the companion could fill its Roche lobe not only at the
end of circularisation, when the companion's radius is maximal, but
also at earlier stages when the radius is smaller but the orbital
separation at periastron is smaller as well.  To find the radius at
any value of $e$, we solve the differential equation,
\begin{equation}\label{diffeq}
\frac{{\rm d}E_\star(R)}{{\rm d}e} = 
    -\frac{{\rm d}E_{\rm orbit}(e)}{{\rm d}e} 
\quad\Rightarrow\quad
-\frac{3Gm^2}{7}\frac{{\rm d}(1/R)}{{\rm d}e} = -\frac{GmM}{a_{\rm c}}e,
\end{equation}
where again assumed conservation of orbital angular momentum.  We find
\begin{eqnarray}\label{overflow}
R(e) = 
  \frac{R_{\rm MS}}{1 + (7 M R_{\rm MS})/(6 m a_{\rm c}) (e^2-e_{\rm ini}^2)}.
\end{eqnarray}
Requiring that this radius is smaller than the Roche limit
\citep[using the approximation of][]{1983ApJ...268..368E} throughout
the circularisation process, we can compute an upper limit to the
initial eccentricity above which Roche lobe overflow occurs.  This
upper limit is also shown in Fig.~\ref{eccini}.

Above, we neglected any energy loss during or after circularisation.
From the isochrones in Fig.~\ref{contour} we conclude that, within the
system age of 10 Myrs, the star cannot have shrunk back significantly
due to radiative losses.
A contraction timescale ($E_{\star}/L$) substantially longer
than the tidal bloating timescale ($E_{\star}/L_{\rm tide}$) is
indeed expected, since the star's luminosity of
$\sim\!0.014~L_\odot$ is substantially smaller than the tidal
luminosity $L_{\rm tide}\ga (e^2_{\rm  ini}GmM/2a_{\rm
c})/\tau=0.08~L_\odot$.
This leaves only one other potential flaw in
our approach: whether the energy is dissipated all through the star.
The dissipation could be confined to a thin layer \citep[see the
discussion in \S5.2 of][]{2000ApJ...529..428V}, and turn the
energy-transport beneath the layer radiative. This would require a
much more sophisticated treatment, if only because of the implications
for the dissipation mechanism itself.  

In summary, if our assumption that the dissipation energy is distributed
throughout the whole star holds, Figure~\ref{ecczero} shows that our
measurement of the companion's temperature and radius is in good
agreement with the initial eccentricity of $\sim\!0.7$ expected from
the formation model.

\subsection{Circularisation timescale\label{taucirc}}

Regardless of the dissipation mechanism, tidal circularisation is
expected to produce a decay of the eccentricity that is exponential.
To reduce the eccentricity from the expected initial eccentricity of
$\sim\!0.7$ to the tiny currently observed value of $\la\!0.005$
therefore requires at least four circularisation timescales.  Since
the system is likely to be in its present state for at most 10 Myr
\citep{2000ApJ...529..428V}, the circularisation timescale must have
been $\la\!2.5~$Myr.

A star of a mass in the observed range is expected to be fully convective.  The
dominant dissipation mechanism in a convective star is thought to be the
friction between the convective motions and the tidal flow, produced by
turbulent viscosity \citep[see, e.g.,][]{1989A&A...220..112Z}. This mechanism
becomes less efficient if -- like in our case -- the convective turnover
timescale is longer than the tidal period, but it is not clear to what extent.
We use two prescriptions for the effect of the timescale mismatch:
\citet{1989A&A...220..112Z}, in which the efficiency is assumed to decrease
linearly with the ratio convective-to-orbital timescale, and
\citet{1997ApJ...486..403G}, in which the efficiency decreases almost
quadratically.  For details on these calculations, see
\citet{2000ApJ...529..428V}.
We should stress that at present it is not clear that either
prescription is reliable; see \citet{1997ApJ...486..403G} for a
discussion.

The resulting theoretical circularisation timescales are listed in
Table~\ref{constab}. It is clear that in \citet{1997ApJ...486..403G}'s
prescription the derived timescales are too long by at least an order
of magnitude, whereas in \citet{1989A&A...220..112Z}'s prescription a
timescale is derived that is consistent with the upper limit cited
above.  However, this is only true if the companion is bloated.  This
reinforces the conclusion derived from studies of the circularisation
of main-sequence binaries that our current theories underestimate the
efficiency of tidal energy dissipation (see
\citealt{2004astro.ph.12147M} for a recent summary of the
observations).

\subsection{The \emph{AIC}-scenario\label{aic}}

The main alternative formation scenario assumes that the system formed
through the accretion-induced collapse (AIC) of a white dwarf.  In
this scenario, the companion was filling its Roche-lobe at the time of
collapse.  The kick, that is required to bring the system to its
present location far outside the cluster core, increases the orbital
separation by only a factor of $\sim1.25$.  Therefore, the companion
should still be close to filling its Roche-lobe.  The only possible
way to explain the observed faint companion, is that the companion was
bloated before collapse, by irradiation from the accreting white dwarf
\citep[as suggested by][]{1993ApJ...415L.115W}.  
As it is not clear whether this is possible by irradiating only 
one side of the star \citep{1996ApJ...467..761K}, we regard the
AIC-scenario as the less likely formation model for this system.

In this scenario, an initial eccentricity of $\sim\!0.2$ is
expected \citep{2000ApJ...529..428V}.  Therefore, the
circularization timescale required to explain the current
circular orbit is roughly the same as for scenario in which the
system is formed by a triple interaction ($\la\!2.5$~Myr).  Even
taking into account that the companion will be close to
Roche-lobe filling, one sees from Table~~\ref{constab} that the
circularization timescale can be matched theoretically only using
the prescription of \citet{1989A&A...220..112Z}.  Thus, our
conclusions in \S\ref{taucirc} remain unchanged.

It is unclear in what precise state the companion is expected to be
immediately after collapse.  For a crude estimate, we assume that at
the moment of collapse, the companion had a radius equal to the Roche
limit, and use Eq.~\ref{eexchange} to compute the critical
eccentricity, substituting $R_{\rm c}=R_{\rm L}=(1/1.25)R_{\rm ini}$.
This yields,
\begin{equation} 
e_{\rm crit}^2 = 0.1383 \left(\frac{m}{0.2M_\odot}\right),
\end{equation}
where we used the same approximations for the semi-major axis and the
Roche limit as in the derivation of Eq.~\ref{ecczero}.  This is
consistent with our measurement of a companion that is bloated to a
radius smaller than the Roche limit, and thus does not add further
constraints on which formation scenario is the correct one.

\section{Conclusion}
We have used VLT observations to measure the temperature and radius of
the likely companion of \mbox{PSR B1718$-$19.} These, in conjunction
with the companion mass inferred from radius and temperature employing
pre-main-sequence cooling tracks, point to a companion that is
bloated, but does not fill its Roche-lobe. 

We have shown that a bloated companion is expected in the context of
the scenario in which the system formed in a a three-body interaction
in the cluster core less than 10 Myrs ago.  With the eccentricity that
is expected to arise from such an event, tidal circularisation can
supply the energy needed for a main sequence star of the observed
mass to expand to the observed radius.  

We tried to compare the observed upper limit on the circularisation
time scale with theory, but were hampered by the uncertainty in the
extent to which the circularisation efficiency is suppressed by the
orbital period being shorter than the convective turnover time.  The
system constitutes an observational constraint on any theoretical
account of tidal circularisation.

The constraints on the properties of the companion could be improved
by even deeper observations, as these would allow a comparison of the
companion with main sequence stars of the same color, thus removing
the uncertainty in distance and reddening.  At this point, however, it
seems more urgent to establish a satisfying theoretical account of the
tidal dissipation mechanism that is dominant in a fully convective
star.

\begin{acknowledgements} 
We thank Ferdi Hulleman and Cees Bassa for help with the photometry, 
and Cees Bassa, Peter Stetson, and Andy Dolphin for useful discussions
on HST photometry.  This research made extensive use of the ADS, SIMBAD, 
and Vizier data bases.
\end{acknowledgements}

\bibliographystyle{aa}
\bibliography{thom}

\begin{thebibliography}{41}
\expandafter\ifx\csname natexlab\endcsname\relax\def\natexlab#1{#1}\fi

\bibitem[{{Amico} {et~al.}(2001){Amico}, {Cuby}, {Devillard}, Jung, \&
  {Lidman}}]{isaac-drg}
{Amico}, P., {Cuby}, J.~G., {Devillard}, N., Jung, Y., \& {Lidman}, C. 2001,
  {ISAAC-SW Data Reduction Guide 1.4}

\bibitem[{{Baggett}(2002)}]{hstdata}
{Baggett}, S., e.~a. 2002, in "HST WFPC2 Data Handbook", Mobasher, B., et al.
  (eds)

\bibitem[{{Baraffe} {et~al.}(1997){Baraffe}, {Chabrier}, {Allard}, \&
  {Hauschildt}}]{1997A&A...327.1054B}
{Baraffe}, I., {Chabrier}, G., {Allard}, F., \& {Hauschildt}, P.~H. 1997, \aap,
  327, 1054

\bibitem[{{Baraffe} {et~al.}(1998){Baraffe}, {Chabrier}, {Allard}, \&
  {Hauschildt}}]{1998A&A...337..403B}
{Baraffe}, I., {Chabrier}, G., {Allard}, F., \& {Hauschildt}, P.~H. 1998, \aap,
  337, 403

\bibitem[{{Bertone} {et~al.}(2004){Bertone}, {Buzzoni}, {Chavez}, \&
  {Rodriguez-Merino}}]{astroph..0406215}
{Bertone}, E., {Buzzoni}, A., {Chavez}, M., \& {Rodriguez-Merino}, L.~H. 2004,
  \aj, in press \verb|[astro-ph/0406215]|

\bibitem[{{Burderi} \& {King}(1994)}]{1994ApJ...430L..57B}
{Burderi}, L. \& {King}, A.~R. 1994, \apjl, 430, L57

\bibitem[{{Chandrasekhar}(1939)}]{1939isss.book.....C}
{Chandrasekhar}, S. 1939, {An introduction to the study of stellar structure}
  (Chicago: The University of Chicago press)

\bibitem[{{Devillard}(1997)}]{eclipse}
{Devillard}, N. 1997, The Messenger, 87, 19

\bibitem[{{Dolphin}(2000)}]{2000PASP..112.1383D}
{Dolphin}, A.~E. 2000, \pasp, 112, 1383

\bibitem[{{Eggleton}(1983)}]{1983ApJ...268..368E}
{Eggleton}, P.~P. 1983, \apj, 268, 368

\bibitem[{{Ergma}(1993)}]{1993A&A...273L..38E}
{Ergma}, E. 1993, \aap, 273, L38+

\bibitem[{{Ergma} {et~al.}(1996){Ergma}, {Sarna}, \&
  {Giersz}}]{1996A&A...307..768E}
{Ergma}, E., {Sarna}, M.~J., \& {Giersz}, M. 1996, \aap, 307, 768

\bibitem[{{Goodman} \& {Oh}(1997)}]{1997ApJ...486..403G}
{Goodman}, J. \& {Oh}, S.~P. 1997, \apj, 486, 403

\bibitem[{{Gu} {et~al.}(2004){Gu}, {Bodenheimer}, \&
  {Lin}}]{2004ApJ...608.1076G}
{Gu}, P., {Bodenheimer}, P.~H., \& {Lin}, D.~N.~C. 2004, \apj, 608, 1076

\bibitem[{{Harris}(1996)}]{1996AJ....112.1487H}
{Harris}, W.~E. 1996, \aj, 112, 1487

\bibitem[{{Heitsch} \& {Richtler}(1999)}]{1999A&A...347..455H}
{Heitsch}, F. \& {Richtler}, T. 1999, \aap, 347, 455

\bibitem[{{Ivanov} \& {Papaloizou}(2004)}]{2004MNRAS.347..437I}
{Ivanov}, P.~B. \& {Papaloizou}, J.~C.~B. 2004, \mnras, 347, 437

\bibitem[{{King} {et~al.}(1996){King}, {Frank}, {Kolb}, \&
  {Ritter}}]{1996ApJ...467..761K}
{King}, A.~R., {Frank}, J., {Kolb}, U., \& {Ritter}, H. 1996, \apj, 467, 761

\bibitem[{{Lyne} {et~al.}(1993){Lyne}, {Biggs}, {Harrison}, \&
  {Bailes}}]{1993Natur.361...47L}
{Lyne}, A.~G., {Biggs}, J.~D., {Harrison}, P.~A., \& {Bailes}, M. 1993, \nat,
  361, 47

\bibitem[{{Mathieu} {et~al.}(2004){Mathieu}, {Meibom}, \&
  {Dolan}}]{2004ApJ...602L.121M}
{Mathieu}, R.~D., {Meibom}, S., \& {Dolan}, C.~J. 2004, \apjl, 602, L121

\bibitem[{{Meibom} \& {Mathieu}(2004)}]{2004astro.ph.12147M}
{Meibom}, S. \& {Mathieu}, R.~D. 2004, \apj, in press

\bibitem[{{Moorwood} {et~al.}(1998){Moorwood}, {Cuby}, {Biereichel}, {Brynnel},
  {Delabre}, {Devillard}, {van Dijsseldonk}, {Finger}, {Gemperlein},
  {Gilmozzi}, {Herlin}, {Huster}, {Knudstrup}, {Lidman}, {Lizon}, {Mehrgan},
  {Meyer}, {Nicolini}, {Petr}, {Spyromilio}, \&
  {Stegmeier}}]{1998Msngr..94....7M}
{Moorwood}, A., {Cuby}, J.-G., {Biereichel}, P., {et~al.} 1998, The Messenger,
  94, 7

\bibitem[{{Paczy{\' n}ski}(1971)}]{1971ARA&A...9..183P}
{Paczy{\' n}ski}, B. 1971, \araa, 9, 183

\bibitem[{{Persson} {et~al.}(1998){Persson}, {Murphy}, {Krzeminski}, {Roth}, \&
  {Rieke}}]{1998AJ....116.2475P}
{Persson}, S.~E., {Murphy}, D.~C., {Krzeminski}, W., {Roth}, M., \& {Rieke},
  M.~J. 1998, \aj, 116, 2475

\bibitem[{{Phinney}(1992)}]{1992RSPTA.341...39P}
{Phinney}, E.~S. 1992, Phil. Trans. R. Soc. London, A, 341, 39

\bibitem[{{Podsiadlowski}(1996)}]{1996MNRAS.279.1104P}
{Podsiadlowski}, P. 1996, \mnras, 279, 1104

\bibitem[{{Ray} {et~al.}(1987){Ray}, {Kembhavi}, \&
  {Antia}}]{1987A&A...184..164R}
{Ray}, A., {Kembhavi}, A.~K., \& {Antia}, H.~M. 1987, \aap, 184, 164

\bibitem[{{Reid} \& {Majewski}(1993)}]{1993ApJ...409..635R}
{Reid}, N. \& {Majewski}, S.~R. 1993, \apj, 409, 635

\bibitem[{{Schlegel} {et~al.}(1998){Schlegel}, {Finkbeiner}, \&
  {Davis}}]{1998ApJ...500..525S}
{Schlegel}, D.~J., {Finkbeiner}, D.~P., \& {Davis}, M. 1998, \apj, 500, 525

\bibitem[{{Seifert} {et~al.}(2000){Seifert}, {Appenzeller}, {Fuertig}, {Stahl},
  {Sutorius}, {Xu}, {Gaessler}, {Haefner}, {Hess}, {Hummel}, {Mantel}, {Meisl},
  {Muschielok}, {Tarantik}, {Nicklas}, {Rupprecht}, {Cumani}, {Szeifert}, \&
  {Spyromilio}}]{2000SPIE.4008...96S}
{Seifert}, W., {Appenzeller}, I., {Fuertig}, W., {et~al.} 2000, in Proc. SPIE,
  M.~Iye, \& A.~F. Moorwood (eds), 4008, 96

\bibitem[{{Stetson}(1987)}]{1987PASP...99..191S}
{Stetson}, P.~B. 1987, \pasp, 99, 191

\bibitem[{{Stetson}(2000)}]{2000PASP..112..925S}
{Stetson}, P.~B. 2000, \pasp, 112, 925

\bibitem[{{Thorsett} \& {Chakrabarty}(1999)}]{1999ApJ...512..288T}
{Thorsett}, S.~E. \& {Chakrabarty}, D. 1999, \apj, 512, 288

\bibitem[{{Van Kerkwijk} {et~al.}(2000){Van Kerkwijk}, {Kaspi}, {Klemola},
  {Kulkarni}, {Lyne}, \& {Van Buren}}]{2000ApJ...529..428V}
{Van Kerkwijk}, M.~H., {Kaspi}, V.~M., {Klemola}, A.~R., {et~al.} 2000, \apj,
  529, 428

\bibitem[{{van Kerkwijk} \& {Kulkarni}(2001)}]{2001A&A...378..986V}
{van Kerkwijk}, M.~H. \& {Kulkarni}, S.~R. 2001, \aap, 378, 986

\bibitem[{{Verbunt}(1994)}]{1994A&A...285L..21V}
{Verbunt}, F. 1994, \aap, 285, L21

\bibitem[{{Verbunt} \& {Phinney}(1995)}]{1995A&A...296..709V}
{Verbunt}, F. \& {Phinney}, E.~S. 1995, \aap, 296, 709

\bibitem[{{Wijers} \& {Paczynski}(1993)}]{1993ApJ...415L.115W}
{Wijers}, R.~A.~M.~J. \& {Paczynski}, B. 1993, \apjl, 415, L115+

\bibitem[{{Wu}(2003)}]{2003sfre.conf..213W}
{Wu}, Y. 2003, in ASP Conf. Ser. 294, 213

\bibitem[{{Zahn}(1989)}]{1989A&A...220..112Z}
{Zahn}, J.-P. 1989, \aap, 220, 112

\bibitem[{{Zwitter}(1993)}]{1993MNRAS.264L...3Z}
{Zwitter}, T. 1993, \mnras, 264, L3

\end{thebibliography}

\end{document}